\documentstyle[aps,12pt,epsf]{revtex}
\def\la{\mathrel{\mathpalette\fun <}}
\def\ga{\mathrel{\mathpalette\fun >}}
\def\fun#1#2{\lower3.6pt\vbox{\baselineskip0pt\lineskip.9pt
  \ialign{$\mathsurround=0pt#1\hfil##\hfil$\crcr#2\crcr\sim\crcr}}}
\begin{document}

\draft
\title{A New Dark Matter Candidate: Non-thermal Sterile Neutrinos}

\author{Xiangdong~Shi and George~M.~Fuller}
\address{Department of Physics, University of California,
San Diego, La Jolla, California 92093-0319}

\date{December 16, 1998}

\maketitle

\begin{abstract}
We propose a new and unique dark matter candidate:
$\sim 100$ eV to $\sim 10$ keV sterile neutrinos
produced via lepton number-driven resonant MSW
(Mikheyev-Smirnov-Wolfenstein) conversion of 
active neutrinos. The requisite lepton number asymmetries
in any of the active neutrino flavors range from 10$^{-3}$
to 10$^{-1}$ of the photon number - well within primordial
nucleosynthesis bounds. The unique feature here is that
the adiabaticity condition of the resonance strongly favors
the production of lower energy sterile neutrinos. The resulting
non-thermal (cold) energy spectrum can cause these sterile
neutrinos to revert to non-relativistic kinematics at an early epoch,
so that free-streaming lengths at or below the dwarf galaxy scale
are possible. Therefore, the main problem associated with light
neutrino dark matter candidates can be circumvented in our model.
\end{abstract}
\bigskip


\newpage
Ample evidence shows that the dominant component of the matter
content of the universe is non-baryonic Dark Matter contributing
a fraction of the critical density $\Omega_{\rm m}\sim 0.2$ to 1
\cite{Kolb,Peebles}. In this letter we suggest a novel means to
allow sterile neutrinos with masses $\sim 100$ eV to $\sim 10$ keV
to be the Dark Matter. We note that the existence of light sterile
neutrinos may be an inevitable conclusion if the neutrino
oscillation interpretations of the Super Kamiokande atmospheric
neutrino data, solar neutrino data and the data from the Los Alamos
Liquid Scintillator Neutrino Detector are correct\cite{Caldwell}.

We envision production of our sterile neutrinos from
active neutrinos via a net lepton number-driven MSW
(Mikheyev-Smirnov-Wolfenstein\cite{MSW})
resonant conversion process occuring near or during
the Big Bang Nucleosynthesis (BBN) epoch.
Only neutrinos which evolve through resonances
adiabatically are efficiently converted.  In turn, lower energy
neutrinos have larger ratios of resonance width to
oscillation length and, hence, evolve more adiabatically.
Therefore, the resonant process can produce a final
energy distribution for the sterile neutrinos which is
grossly non-thermal and skewed to very low energies.
These neutrinos have energy distribution functions
which are \lq\lq cold\rq\rq . This, coupled with their
relatively large rest mass, implies that they go
non-relativistic at quite early epochs, therefore
essentially becoming Cold Dark Matter (CDM), albeit
with a cut-off at or below the dwarf galaxy scale. Below
this cut-off scale, free streaming of these neutrinos can
effectively damp out density fluctuations. Since this
cut-off scale is much lower than any previous light
neutrino dark matter models (e.g., Hot Dark Matter, HDM\cite{HDM},
Warm Dark Matter, WDM\cite{WDM}, and mixed Cold plus Hot CHDM\cite{CHDM}
models), we deem these sterile neutrinos to be \lq\lq
Cool Dark Matter\rq\rq\  (CoolDM). 

The contribution to the matter density of our universe
from an active neutrino species ($\nu_e$, $\nu_\mu$ or
$\nu_\tau$) with mass $m_\nu$ is
\begin{equation}
\Omega_\nu\approx {m_\nu\over 91.5h^2\,{\rm eV}}.
\end{equation}
Here $h\equiv H_0/(100\,{\rm km/sec/Mpc})$
where $H_0$ is the Hubble constant. Therefore, active neutrinos
with masses in the ${\cal O}(10)$ eV range naturally serve as
Dark Matter candidates.
They are \lq\lq hot\rq\rq\  because their kinetic temperature is 
relatively high as a result of their light masses.
It has been shown that models with HDM as the sole
Dark Matter ingredient of the universe fail to form
galaxies at an early enough epoch. This is due to the large
free streaming of light neutrinos which damp out density fluctuations
below the free streaming scale \cite{HDM}
\begin{equation}
\lambda_{\rm f.s.}\sim 40\left({m_\nu\over 30\,{\rm eV}}\right)^{-1}\,
{\rm Mpc}.
\label{free}
\end{equation}
This generic problem of the large free streaming length of light
neutrinos is difficult to circumvent. One possibility is to have 
non-gaussian \lq\lq seeds\rq\rq\  in the primordial density 
fields\cite{SHDM}.  Another is to have light neutrinos comprise
only a small fraction of the dark matter 
($\Omega_\nu/\Omega_{\rm m}\la 0.2$) \cite{CHDM,Primack}.

WDM models consisting of ${\cal O}(10^2)$ eV sterile neutrinos
have a free-streaming length much shorter than that of
HDM\cite{Colombi}. Because of this free-streaming,
WDM models have a reduced amplitude of density fluctuations at
the galaxy cluster scale, which fits observations of structures
at this scale better than the {\sl COBE}-normalized
standard CDM (sCDM) model (with $\Omega_{\rm m}=1$ and $h=0.5$)
\cite{Colombi}. However, density fluctuations at smaller galaxy scales
($\sim 1h^{-1}$ Mpc) in WDM models may be too small \cite{Colombi}
to accomodate observations of Damped Lyman-$\alpha$ Systems at high
redshifts, and observations of \lq\lq lyman-break\rq\rq\  galaxies 
which are highly-clustered galaxies at redshifts $z\sim 3$ \cite{Steidel}.

The neutrino Cool Dark Matter candidate proposed here is
\lq\lq colder\rq\rq\ (i.e., with a much smaller free-streaming
length) than all three classes of models mentioned above.
In particular, the cold non-thermal energy distribution of our
candidate sterile neutrinos causes them to move much more
slowly than WDM sterile neutrinos with similar masses. The
small free streaming lengths in CoolDM models enable a
circumvention of the generic problem associated with light
neutrino dark matter candidates. 

The production mechanism of our sterile neutrino
CoolDM candidate is very different from that of its
closest kin - the sterile neutrino WDM candidate.
In the model proposed by Dodelson and Widrow \cite{WDM},
sterile neutrinos were produced by active-sterile neutrino
oscillation at the BBN epoch.  The oscillation was driven
by finite-temperature matter effects\cite{Mattereffect}.
The end-result was a sterile neutrino population with an
energy spectrum resembling that of an active neutrino
species but with a overall suppression in normalization
\cite{WDM}. The average energy of these sterile neutrinos
is therefore the same as the active ones, $\approx 3.151T$,
where $T$ is the temperature of the active neutrino species.
In our model, the sterile neutrinos are produced by a {\bf resonant}
active-sterile neutrino transformation driven by a pre-existing
lepton number asymmetry.  The resonant transformation is only 
adiabatic at the low energy portion of the neutrino energy spectrum.
This feature creates a non-thermal sterile neutrino population, 
with only the low energy states populated.

We quantify this production mechanism in the density
matrix formalism\cite{Shi}.  In this formalism, the
$\nu_\alpha\leftrightarrow\nu_s$ ($\alpha = e$, $\mu$
or $\tau$, and $\nu_s$ is the sterile neutrino) system
is described by a four vector ($P_0$, {\bf P}). The
number densities of the neutrinos are
$n_{\nu_\alpha}\equiv (P_0+P_z)/2$ and
$n_{\nu_s}\equiv (P_0-P_z)/2$. The evolution of the 
mixing system in the BBN epoch satisfies the equation:
\begin{equation}
\dot {\bf P}={\bf V}\times{\bf P}+\dot P_0{\bf\hat z}-D{\bf P_\bot},
\label{master}
\end{equation}
where {\bf P}$_\bot=P_x{\bf\hat x}+P_y{\bf\hat y}$. The $D$ term
is the quantum damping term that acts to reduce the mixing system
into flavor eigenstates and suppress the oscillation.
Numerically $D\sim G_F^2T^5$, where $G_F$ is the Fermi constant.
The vector {\bf V} is the effective potential of the oscillation.
Its components are
\begin{equation}
V_x={\delta m^2\over 2E}\sin 2\theta,\quad
V_y=0,\quad
V_z=-{\delta m^2\over 2E}\cos 2\theta+V_\alpha^L+V_\alpha^T,
\label{potential}
\end{equation}
where $\delta m^2\equiv m_{\nu_s}^2-m_{\nu_\alpha}^2\approx m_{\nu_s}^2$,
$\theta$ is the vacuum mixing angle, and $E$ is neutrino energy.
The matter-antimatter asymmetry contribution to the effective potential
is\cite{Mattereffect}
\begin{equation}
V_\alpha^L\approx 0.35 G_FT^3\Bigl[L_0+2L_{\nu_\alpha}+
                  \sum_{\beta\ne\alpha}L_{\nu_\beta}\Bigr]
\label{VL}
\end{equation}
where $L_0$ represents the contributions from the baryonic 
asymmetry and electron-positron asymmetry, $\sim 10^{-10}$;
and $L_{\nu_\beta}$ is the asymmetry in the other active 
neutrino species $\nu_\beta$. For convenience, we will 
denote ${\cal L}\equiv 2L_{\nu_\alpha}+\sum_{\alpha\ne\beta}L_{\nu_\beta}$.
The contribution to {\bf V} from a thermal neutrino background
is $V_\alpha^T \sim -10^2G_F^2ET^4$ \cite{Mattereffect}.

Similar quantities for the anti-neutrino $\bar\nu_\alpha\leftrightarrow
\bar\nu_s$ can be defined and they satisfy
\begin{equation}
\dot {\bf\bar P}={\bf\bar V}\times{\bf\bar P}+\dot{\bar P_0}
{\bf\hat z}-{\bar D}{\bf\bar P_\bot}.
\label{antimaster}
\end{equation}
where $\bar V_x=V_x,\,\bar V_y=V_y,\,\bar V_z
=-\delta m^2\cos 2\theta/2E-V_\alpha^L+V_\alpha^T$, and $\bar D\approx D$.

The sign of ${\cal L}$ does not matter in the Dark
Matter problem.  For definitiveness we assume that
it is positive, so that $V_\alpha^L$ is positive. 
Then the $\nu_\alpha\leftrightarrow\nu_s$ oscillation
system encounters a resonance ($V_z=0$, or equivalently,
a maximally matter-enhanced mixing) as a result of the 
non-zero ${\cal L}$ at a temperature
\begin{equation}
T_{\rm res}\approx 9\left({m_{\nu_s}\over 10^2\,{\rm eV}}\right)^{1/2}
	\left({{\cal L}\over 0.1}\right)^{-1/4}
	\epsilon^{-1/4}\,{\rm MeV},
\label{Tres}
\end{equation}
where $\epsilon\equiv E/T$. At $T\sim T_{\rm res}$,
$V_\alpha^T$ is completely negligible. We have implicitly
taken $\cos\theta\approx 1$ since the mixing angle must be
small.  The anti-neutrino $\bar\nu_\alpha\leftrightarrow
\bar\nu_s$ oscillation, on the other hand, has $\vert V_z\vert >
\vert -\delta m^2/2E\vert$. It is therefore suppressed relative 
to a vacuum oscillation situation.
The net effect is destruction of $\cal L$,
because $\nu_\alpha$ is resonantly transformed
into sterile neutrinos while $\bar\nu_\alpha$ is not. 

Quite similar to the putative matter-enhanced resonance
transition in the solar neutrino problem, the
adiabaticity condition of the $\nu_\alpha\rightarrow\nu_s$
resonant transformation at the resonant energy bin
$\epsilon_{\rm res}$ is
\begin{equation}
V_x^2\left\vert{{\rm d}\epsilon_{\rm res}\over {\rm d}V_z}\right\vert
\left\vert{{\rm d}\epsilon_{\rm res}\over {\rm d}t}\right\vert^{-1}>1.
\label{adiabaticity}
\end{equation}
In the equation, $\vert V_x\vert$ is the transformation rate,
$\vert V_x({\rm d}\epsilon_{\rm res}/{\rm d}V_z)\vert$
is the energy width of the resonance, and
$\vert {\rm d}\epsilon_{\rm res}/{\rm d}t\vert$ is
the the speed of movement of the resonance energy bin
across the neutrino spectrum as $\nu_\alpha$ neutrinos
of different energies encounter resonance at different
temperatures. (Strictly speaking, our expression for the
rate and the energy width of the resonances are only
valid when collisions are not important, i.e., $D\ll \vert V_x\vert$.
However, the adiabaticity condition is nonetheless the
same when $D\gg \vert V_x\vert$ because the suppression
in rate due to collisions is compensated by the increase
in the energy width of the resonance.) From
Eq.~(\ref{potential}) we have
\begin{equation}
\left\vert V_x{{\rm d}\epsilon_{\rm res}\over {\rm d}V_z}\right\vert
\approx \epsilon_{\rm res}\sin 2\theta.
\end{equation}
From Eq.~(\ref{Tres}), we have
\begin{equation}
{{\rm d}\epsilon_{\rm res}\over {\rm d}t}
\approx \epsilon_{\rm res}\left[4H-
{{\rm d}{\cal L}/{\rm d}t
\over {\cal L}}\right],
\label{speed}
\end{equation}
where $H\approx 5.5T^2/m_{\rm pl}$ is the Hubble expansion rate
and $m_{\rm pl}\approx 1.22\times 10^{28}$ eV is the Planck mass.
The adiabaticity condition is therefore satisfied if
\begin{equation}
4\times 10^9\left({m_{\nu_s}\over 10^2\,{\rm eV}}\right)^{1/2}
\left({{\cal L}\over 0.1}\right)^{3/4}\epsilon_{\rm res}^{-1/4}
\sin^2 2\theta\left[{1\over 1-({\rm d}{\cal L}/{\rm d}t)/4H{\cal L}}
\right]>1.
\label{adiabatic}
\end{equation}
Assuming adiabatic neutrino evolution through resonances,
we have ${\rm d}{\cal L}/{\rm d}t
=f(\epsilon_{\rm res}){\rm d}{\epsilon_{\rm res}}/{\rm d}t
=f(\epsilon_{\rm res})\epsilon_{\rm res}[4H-
({\rm d}{\cal L}/{\rm d}t)/{\cal L}]$ where $f(\epsilon)$
is the neutrino distribution function.  We therefore derive
${\cal L}={\cal L}^{\rm init}-\int_0^{\epsilon_{\rm res}}
f(\epsilon){\rm d}\epsilon$ where superscript \lq\lq init\rq\rq\
indicates initial values.  As a result,
$\vert ({\rm d}{\cal L}/{\rm d}t)/{\cal L}\vert\la H$ is always true
unless $\vert {\cal L}\vert\ll \vert {\cal L}^{\rm init}\vert$.
This implies that the adiabaticity condition Eq.~(\ref{adiabatic})
holds true for $\nu_\alpha\leftrightarrow\nu_s$ vacuum mixing that
is not too small ($\sin^22\theta\ga 10^{-9}$), until most of
$\cal L$ is destroyed by the resonant $\nu_\alpha\leftrightarrow\nu_s$
transformation. The final $\cal L$ is $\sim 0$ after the resonant
conversion process.  In this limit, the total change of the 
$\nu_\alpha\bar\nu_\alpha$ asymmetry is
$\Delta L_{\nu_\alpha}\approx {\cal L}^{\rm init}/2$.
This change is entirely due to the $\nu_\alpha$ to 
$\nu_s$ transformation, implying that the $\nu_s$ sea
produced has a number density that is a fraction
\begin{equation}
F\approx {4\over 3}\Delta L_{\nu_\alpha}
\end{equation}
of the number density of an active neutrino species.
The sterile neutrino contribution to the matter density today is then
\begin{equation}
\Omega_{\nu}\approx F\left({m_{\nu_s}\over 91.5h^2\,{\rm eV}}\right)
\approx \left({m_{\nu_s}\over 343\,{\rm eV}}\right)
\left({h\over 0.5}\right)^{-2}
\left({2L_{\nu_\alpha}+\sum_{\beta\ne\alpha}L_{\nu_\beta}\over 0.1}\right),
\label{cdm}
\end{equation}
with $\alpha ,\,\beta=e,\,\mu,\,\tau$.
Here we have assumed that all neutrino
asymmetries are their initial values 
and we have dropped the superscript 
\lq\lq init\rq\rq\ ($L_{\nu_\beta}$
will not change anyway).

From Eq.~(\ref{potential}) and (\ref{cdm}) we note that if
the sign of $2L_{\nu_\alpha}+\sum_{\beta\ne\alpha}L_{\nu_\beta}$
is negative, the anti-neutrino counterpart of $\nu_s$ will be
resonantly produced from $\bar\nu_\alpha$, and may therefore
become the Dark Matter candidate if the counterpart of Eq.~(\ref{cdm})
is satisfied.

As the transition is only adiabatic when ${\cal L}$
is a significant fraction of its initial value,
only the low energy sterile neutrinos (which encounter resonances
first while ${\cal L}$ is still relatively large, see
Eq.~[\ref{Tres}]) are resonantly produced.  Higher energy
neutrinos go through the resonance later, when most of
${\cal L}$ is damped. Evolution through resonances
for these high energy neutrinos is therefore non-adiabatic,
producing no significant $\nu_\alpha\rightarrow\nu_s$
conversion.  The end result is a sterile
neutrino energy spectrum as illustrated in Figure 1, with a
characteristic mean neutrino energy $E$ approximately satisfying
$\int_0^{E/T}f(E^\prime/T){\rm d}
(E^\prime/T)\approx 8\Delta L_{\nu_\alpha}/3\approx 4{\cal L}/3$.
As an example, for ${\cal L}=0.02$, we have $E/T$ is 0.7, i.e.,
%As an example, for ${\cal L}=0.01$, this characteristic $E/T$ is 0.6,
less than one-quarter of the average $E/T$ for active species.
The comoving free-streaming length of these sterile neutrinos becomes
\begin{equation}
\lambda_{\rm f.s.}\sim 40\left({m_\nu\over 30\,{\rm eV}}\right)^{-1}
\left({E/T\over 3.15}\right)\,{\rm Mpc},
\end{equation}
which for a given sterile neutrino mass can be more than a factor of several
shorter than that calculated from Eq.~(\ref{free}).

In this letter we limit our discussion of the sterile neutrino
production to a temperature range below the Quark-Hardron
phase transition, $\sim 150$ MeV.  The reason for this
is that the neutrino transformation process can be modeled
confidently below this temperature. With this restriction,
Eq.~(\ref{Tres}) shows that the pre-existing
${\cal L}$ has to be $\ga 10^{-3}$. The mass
of the sterile neutrino Dark Matter candidate is then $\la 10$
keV from  Eq.~(\ref{cdm}). (${\cal L}$ and the sterile neutrino
mass are limited simultaneously because their product must
yield $\Omega_\nu\sim 1$.)
If we also restrict the sterile neutrino mass to be
$m_{\nu_s}\ga 10^2$ eV in order to have potentially 
successful structure formation at galaxy scales, 
Eq.~(\ref{cdm}) implies that ${\cal L}\la 10^{-1}$.
For example, if the pre-existing $2L_{\nu_\mu}+L_{\nu_e}+L_{\nu_\tau}$
is 0.02, a sterile neutrino with $m_{\nu_s}=1160$ eV and mixing with
$\nu_\mu$ will yield $\Omega_\nu=0.4$ for $h=0.65$.
The comoving free-streaming length of this Dark Matter
is only $\lambda_{\rm f.s.}\sim 0.4$ Mpc, giving rise to
a mass-scale cut-off of $M\sim 10^{10}M_\odot$, roughly
the scale of dwarf galaxies.  In this case, structure 
formation in CoolDM models differs from that in CDM models
only at sub-galactic scales, such as in halo structure and
formation history at high redshift. These differences may 
potentially be tested by observations of galaxy rotation 
curves \cite{Silk} and high redshift galaxies\cite{Steidel,Wolfe}.
However, if the mass of the sterile neutrinos is as high as 10 
keV, resulting from ${\cal L}\sim 10^{-3}$, then CoolDM essentially
becomes CDM. It is intriguing that oscillation between active
neutrinos and keV sterile neutrinos in supernovae in the presence
of strong magnetic fields may potentially give rise to \lq\lq 
pulsar kicks\rq\rq\ \cite{Kusenko}.

An ${\cal L}$ between $\sim 10^{-3}$ and $\sim 10^{-1}$
necessarily implies that the asymmetries in one or more
active neutrino species are of the same order (or larger
if there is cancellation between asymmetries).
Lepton asymmetries between $\sim 10^{-3}$ and
$\sim 10^{-1}$ are not ruled out by Big Bang 
Nucleosynthesis \cite{BBNbound}. In fact, if
this asymmetry resides in the $\nu_\mu$ or $\nu_\tau$ sector,
its impact on the primordial $^4$He yield (which gives the
constraint on lepton asymmetry) is well below the current
detection level.  If some of the asymmetry is in the $\nu_e$
sector, with a magnitude $\ga 10^{-2}$, its impact on the
primordial $^4$He yield is quite appreciable. 
Just where the limit on $L_{\nu_e}$ stands is hard to quantify precisely
because of the large uncertainty in the primordial $^4$He abundance
measurements\cite{Olive,Thuan}.
In particular, Kohri {\sl et al.}\cite{Kohri} has 
argued that an $L_{\nu_e}\approx 0.015$ is what
is needed to bring a low set of primordial $^4$He
abundance values \cite{Olive} into agreement with the 
measured low primordial deuterium abundance\cite{Tytler}. 

On the other hand, in individual neutrino sectors,
lepton asymmetries of order $\sim 10^{-3}$ to
$\sim 10^{-1}$ are very large compared
to the baryon asymmetry, $\sim 10^{-10}$. But this 
is not to say that their sum, the total lepton number
asymmetry, cannot be as small as $\sim 10^{-10}$.
In the absence of such a cancellation, however,
the resultant large lepton number must be generated
below the Electroweak phase transition to avoid
the transfer between lepton number and baryon number.
Models that generate a large lepton number despite
a small baryon number asymmetry indeed exist\cite{Affleck,FootVolkas}.

In summary, we have proposed that $\sim 10^2$ eV to
$\sim 10$ keV sterile neutrinos $\nu_s$ will be a Cool
Dark Matter candidate if they were produced in the 
early universe via active-sterile neutrino resonant
oscillation in the presence of a lepton number
asymmetry.  The amplitude of this lepton number
asymmetry is $\sim 10^{-3}$ to $\sim 10^{-1}$ for
at least one active neutrino species, which is
consistent with constraints from Big Bang Nucleosynthesis.
The minimal set of free parameters in our Cool Dark Matter model,
and its potentially testable predictions for halo structures
and proto-galaxy formation, all make the Cool Dark Matter
scenario interesting and worthy of further invistigations.

X.~S. and G.~M.~F. are supported by NSF grant PHY98-00980 at UCSD.

\newpage
\noindent{\bf Figure Captions:}

\noindent
Figure 1. Illustration of the energy spectrum of the sterile neutrinos
produced vs. the thermal spectrum of the active neutrino species.
\bigskip

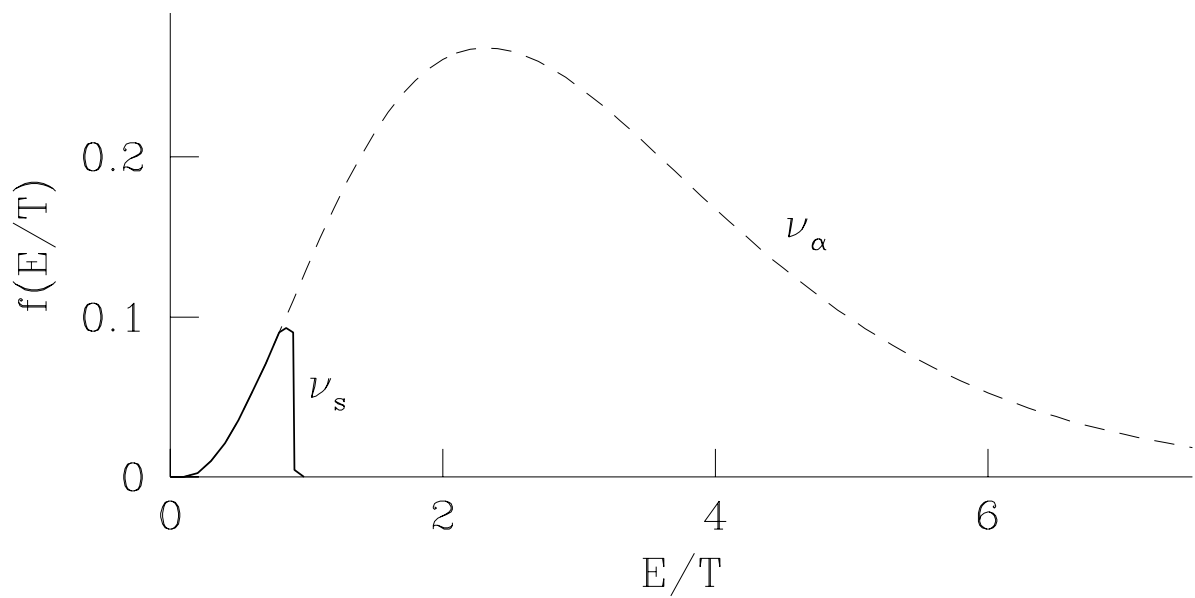
\end{document}